\documentclass[aps,prd,twocolumn,showpacs,floats,floatfix,letterpaper,nofootinbib,superscriptaddress,]{revtex4}

\usepackage{amssymb,amsmath,latexsym,mathrsfs}
\usepackage{graphicx}
\usepackage{epsfig}
\usepackage{multirow}
\usepackage{array}
\usepackage{color}
\usepackage{xspace} 

\newcommand{\bea}{\begin{eqnarray}}
\newcommand{\eea}{\end{eqnarray}}

\newcommand{\Planck}{\textit{Planck}\xspace}
\newcommand{\LCDM}{$\Lambda$CDM\xspace}
\newcommand{\ns}{\ifmmode n_{_{\rm S}} \else $n_{_{\rm S}}$\fi}
\newcommand{\nT}{\ifmmode n_{_{\rm T}} \else $n_{_{\rm T}}$\fi}

\newbox\tablebox    \newdimen\tablewidth
\def\leaderfil{\leaders\hbox to 5pt{\hss.\hss}\hfil}
\def\endPlancktable{\tablewidth=\columnwidth
    $$\hss\copy\tablebox\hss$$
    \vskip-\lastskip\vskip -2pt}

\def\tablenote#1 #2\par{\begingroup \parindent=0.8em
    \abovedisplayshortskip=0pt\belowdisplayshortskip=0pt
    \noindent
    $$\hss\vbox{\hsize\tablewidth \hangindent=\parindent \hangafter=1 \noindent
    \hbox to \parindent{$^#1$\hss}\strut#2\strut\par}\hss$$
    \endgroup}
\def\doubleline{\vskip 3pt\hrule \vskip 1.5pt \hrule \vskip 5pt}

\begin{document}


\title{A comment on power-law inflation with a dark radiation component}

\author{Eleonora Di Valentino}
\affiliation{Institut d'Astrophysique de Paris (UMR7095: CNRS \& UPMC- Sorbonne Universities), F-75014, Paris, France}
\affiliation{Sorbonne Universit\'es, Institut Lagrange de Paris (ILP), F-75014, Paris, France}
\author{Fran\c cois R. Bouchet}
\affiliation{Institut d'Astrophysique de Paris (UMR7095: CNRS \& UPMC- Sorbonne Universities), F-75014, Paris, France}

\begin{abstract}

\noindent Tram et al. 2016 recently pointed out in \cite{Tram:2016rcw} that power-law inflation in presence of a dark radiation component may relieve the $3.3\,\sigma$ tension which exists within standard \LCDM between the determination of the local value of the Hubble constant by Riess et al. (2016) \cite{R16} and the value derived from CMB anisotropy data \cite{planck2015} by the \Planck collaboration.
In this comment, we simply point out that this interesting proposal does not help in solving the $\sigma_8$ tension between the \Planck data and, e.g., the weak lensing measurements. Moreover, when the latest constraints on the reionization optical depth obtained from \Planck HFI data \cite{newtau} are included in the analysis, the $H_0$ tension reappears and this scenario looses appeal. 
\end{abstract}


\maketitle

\section{Introduction}
\label{sec:introduction}

The tension in the Hubble constant between the constraints coming from the \Planck satellite \cite{planck2013} and \cite{planck2015} and the local measurements of Riess at al. \cite{R11} and \cite{R16} has recently gained statistical significance ($3.3\,\sigma$) within the \LCDM framework, especially when considering the new constraints on the reionization optical depth obtained with \Planck HFI data \cite{newtau}.

Many proposals have been suggested to solve this tension
(see for example \cite{planck2013,planck2015,darkradiation,interacting,voids,Ben-Dayan:2014swa,variance,DiValentino:2016hlg,Bernal:2016gxb,Ko:2016uft}).
Two possible extensions to the $\Lambda$CDM scenario have attracted significant attention. One is the  possibility of a dark radiation component with $N_{\rm eff}>3.046$, which does not seem workable any more in light of the new \Planck HFI constraint on the optical depth \cite{newtau} for which $N_{\rm eff}=2.91_{-0.37}^{+0.39}$ at $95\%$ c.l. from Planck TTTEEE+SIMlow. The other extension is a dark energy equation of state different from $w = -1$ (see \cite{DiValentino:2016hlg}). 

However, recently, the authors of \cite{Tram:2016rcw} pointed out that, by considering a power-law inflation (hereafter PLI), introduced the first time by \cite{Lucchin:1984yf}, in presence of a dark radiation component, the local value of the Hubble constant by Riess et al. 2016 \cite{R16} and the most recent CMB anisotropy data by the \Planck collaboration \cite{planck2015} are in perfect agreement, provided $\Delta N_{\rm eff}=0.62\pm0.17$ at $68\%$ c.l..

Given the well-known correlations between parameters within \LCDM, we consider in this comment the implication of this scenario (PLI with a free dark radiation) for the $z=0$ linear power normalisation, $\sigma_8$  or clustering parameter. Indeed, within \Planck-normalised \LCDM, there is already a 2\,$\sigma$ tension with the weak lensing measurements of $\sigma_8$ from the CFHTLenS survey \cite{Heymans:2012gg, Erben:2012zw} and KiDS-450 \cite{Hildebrandt:2016iqg}. 
We also assess the effect of adding the CMB polarization $B$ modes constraint provided by the common analysis of \Planck, BICEP2 and Keck Array \cite{BKP}, or the new determination on the reionization optical depth obtained with \Planck HFI data \cite{newtau} (which was not considered by \cite{Tram:2016rcw}).

\section{Method}

As a baseline, we will explore simultaneously 8 parameters of \LCDM. These include the baryon and cold dark matter energy densities $\Omega_bh^2$ and $\Omega_ch^2$, the ratio between the sound horizon and the angular diameter distance at decoupling $\Theta_{s}$ and the reionization optical depth $\tau$. For the inflationary parameters, we consider the scalar spectral index, \ns, the amplitude of the primordial spectrum, $A_\mathrm{S}$, and a contribution of primordial gravitational waves with a tensor-to-scalar ratio of amplitude $r$ at the pivot scale $k_0=0.05 hMpc^{-1}$. Finally, we also vary the effective number of relativistic degrees of freedom $N_{\rm eff}$. We therefore include $r$ and $N_{\rm eff}$ to the minimum 6 base \LCDM parameters which provide an adequate fit to \Planck data (assuming a flat Universe). These parameters are explored within the range of the conservative priors reported in Table~\ref{priors}.
\begin{table}[htbp]
\begingroup
\newdimen\tblskip \tblskip=5pt
\caption{External priors on the cosmological parameters assumed in this work.}
\label{priors}
\nointerlineskip
\setbox\tablebox=\vbox{
   \newdimen\digitwidth
   \setbox0=\hbox{\rm 0}
   \digitwidth=\wd0
   \catcode`*=\active
   \def*{\kern\digitwidth}
   \newdimen\signwidth
   \setbox0=\hbox{+}
   \signwidth=\wd0
   \catcode`!=\active
   \def!{\kern\signwidth}
\halign{\hbox to 1.0in{#\leaderfil}\tabskip=3em&
        \hfil#\hfil\tabskip=0pt\cr
\noalign{\doubleline}
Parameter                    & Prior\cr
\noalign{\vskip 5pt\hrule\vskip 3pt}
$\Omega_{\rm b} h^2$         & $[0.005,*0.10]$\cr
$\Omega_{\rm cdm} h^2$       & $[0.001,*0.99]$\cr
$\Theta_{\rm s}$             & $[0.5**,10.0*]$\cr
$\tau$                       & $[0.01*,*0.80]$\cr
$n_s$                        & $[0.8**,*1.20]$\cr
$\log[10^{10}A_{s}]$         & $[2.0**,*4.0*]$\cr
$r$                          &  [0.***,*0.50]\cr
$N_{\rm eff}$                &  [2.0**,*5.0*]\cr
\noalign{\vskip 5pt\hrule\vskip 3pt} }}
\endPlancktable
\endgroup
\end{table}
In a second step, we repeat the same analysis considering the PLI model by imposing the following inflation consistency relationship \cite{Martin:2013tda}:
\begin{equation}
r=\frac{16 \nT}{\nT -2} 
\label{eq:PLIdef}
\end{equation}
with
\begin{equation}
\nT=\ns-1 \,.
\end{equation}

\begin{table*}
\begin{center}\footnotesize
\scalebox{1.04}{\begin{tabular}{lccccccc}
\hline \hline
         & Planck TT & Planck  TT & Planck TT& Planck TTTEEE&Planck TTTEEE&Planck TTTEEE\\                     
         & + lowTEB     &        + lowTEB + lensing  & + tau055      &  + lowTEB &+tau055 &+ lowTEB + BKP\\  
\hline
\hspace{1mm}\\

$\Omega_{\textrm{b}}h^2$& $0.02240\,\pm 0.00037 $& $0.02242\,_{-0.00041}^{+0.00035}$    & $0.02182\,\pm0.00034 $& $0.02219\,\pm 0.00025$& $0.02194\,\pm 0.00023$ & $0.02219\,\pm 0.00024$   \\
\hspace{1mm}\\

$\Omega_{\textrm{c}}h^2$& $0.1214\, _{-0.0043}^{+0.0039}$& $0.1203\,\pm 0.0039$    & $0.1178\,\pm0.0041 $& $0.1191\,\pm0.0031$ & $0.1177\,\pm0.0031$ & $0.1191\,\pm 0.0031$   \\
\hspace{1mm}\\

$\tau$& $0.082\,_{-0.025}^{+0.021}$& $0.072\,_{-0.021}^{+0.018}$    & $0.0572\,\pm 0.0089$& $0.076\,\pm 0.018$& $0.0591\,\pm 0.0088$ & $0.077\,\pm 0.018$    \\
\hspace{1mm}\\

$n_S$& $0.974\,\pm 0.016$& $0.976\,_{-0.017}^{+0.014}$    & $0.946\,\pm 0.016 $& $0.963\,\pm 0.010$& $0.9510\,\pm 0.0095$ & $0.9628\,\pm 0.0096$   \\
\hspace{1mm}\\

$log(10^{10}A_S)$& $3.102\,_{-0.052}^{+0.046}$& $3.078\,\pm0.041$    & $3.041\,\pm 0.023$& $3.086\,\pm 0.038$& $3.047\,\pm 0.020$ & $3.088\,\pm 0.037$    \\
\hspace{1mm}\\

$H_0$ &      $68.9\, _{-3.2}^{+2.7}$&      $ 69.3\,_{-3.1}^{+2.6}$ & $ 64.0\,_{-2.9}^{+2.5}$   &  $ 66.9\,\pm 1.7$ &  $ 65.1\,\pm 1.5$  &  $ 66.9\,\pm 1.6$ \\
\hspace{1mm}\\

$\sigma_8$   & $ 0.837\,_{-0.025}^{+0.022}$   & $ 0.824\,_{-0.021}^{+0.018}$   & $ 0.809\,\pm 0.014$ &  $ 0.827\,\pm0.018$ &  $ 0.809\,\pm0.012$ &  $ 0.828\,\pm0.017$  \\
\hspace{1mm}\\

$N_{\rm eff}$ &  $3.23\,_{-0.36}^{+0.30}$ &  $3.22\,_{-0.34}^{+0.29}$  & $2.73\,_{-0.34}^{+0.30}$&  $3.00\,\pm0.21$ &  $2.81\,\pm0.20$ &  $3.00\,\pm0.20$  \\
\hspace{1mm}\\

$r$ &  $<0.0529$ &  $<0.0612$  & $<0.102$&  $<0.0487$ &  $<0.103$& $0.036\,_{-0.035}^{+0.011}$   \\
\hspace{1mm}\\
\hline
\hline

\end{tabular}}
\caption{$68 \% $ c.l. constraints on cosmological parameters in our extended $\Lambda$CDM+r+$N_{\rm eff}$ scenario from different combinations of datasets.}
\label{table1}
\end{center}
\end{table*}

\begin{table*}
\begin{center}\footnotesize
\scalebox{1.04}{\begin{tabular}{lccccccc}
\hline \hline
         & Planck TT & Planck  TT & Planck TT& Planck TTTEEE&Planck TTTEEE&Planck TTTEEE\\                     
         & + lowTEB     &        + lowTEB + lensing  & + tau055      &  + lowTEB &+tau055 &+ lowTEB + BKP\\  
\hline
\hspace{1mm}\\

$\Omega_{\textrm{b}}h^2$& $0.02275\,\pm 0.00024 $& $0.02273\,0.00024$    & $0.02231\,\pm0.00026 $& $0.02262\,\pm 0.00020$& $0.02227\,\pm 0.00019$ & $0.02275\,\pm 0.00016$   \\
\hspace{1mm}\\

$\Omega_{\textrm{c}}h^2$& $0.1240\, \pm0.0036$& $0.1226\,\pm 0.0033$    & $0.1222\,\pm0.0037 $& $0.1231\,\pm0.0028$ & $0.1211\,\pm0.0030$ & $0.1243\,\pm 0.0028$   \\
\hspace{1mm}\\

$\tau$& $0.095\,\pm0.020$& $0.085\,\pm0.015$    & $0.0602\,\pm 0.0088$& $0.092\,\pm 0.017$& $0.0620\,\pm 0.0087$ & $0.100\,\pm 0.017$    \\
\hspace{1mm}\\

$n_S$& $0.9918\,_{-0.0035}^{+0.0072}$& $0.9911\,_{-0.0038}^{+0.0074}$    & $0.9735\,\pm 0.0083 $& $0.9849\,\pm 0.0057$& $0.9695\,\pm 0.0062$ & $0.9907\,_{-0.0034}^{+0.0042}$   \\
\hspace{1mm}\\

$log(10^{10}A_S)$& $3.134\,\pm0.039$& $3.111\,\pm0.028$    & $3.060\,\pm 0.020$& $3.126\,\pm 0.035$& $3.062\,\pm 0.020$ & $3.146\,\pm 0.034$    \\
\hspace{1mm}\\

$H_0$ &      $72.0\, _{-1.1}^{+1.5}$&      $ 72.0\,_{-1.1}^{+1.5}$ & $ 68.7\,\pm1.7$   &  $ 70.3\,\pm 1.1$ &  $ 67.8\,\pm 1.2$  &  $ 71.32\,\pm 0.84$ \\
\hspace{1mm}\\

$\sigma_8$   & $ 0.854\,\pm0.018$   & $ 0.841\,\pm0.012$   & $ 0.822\,\pm 0.013$ &  $ 0.850\,\pm0.017$ &  $ 0.821\,\pm0.011$ &  $ 0.861\,\pm0.016$  \\
\hspace{1mm}\\

$N_{\rm eff}$ &  $3.55\,_{-0.17}^{+0.19}$ &  $3.51\,\pm0.17$  & $3.24\,\pm0.22$&  $3.39\,\pm0.15$ &  $3.14\,\pm0.17$ &  $3.51\,\pm0.13$  \\
\hspace{1mm}\\

$r$ &  $0.065\,_{-0.062}^{+0.020}$ &  $0.071\,_{-0.060}^{+0.027}$  & $0.209\,\pm0.064$&  $0.120\,\pm0.045$ &  $0.240\,\pm0.048$& $0.074\,_{-0.033}^{+0.027}$   \\
\hspace{1mm}\\
\hline
\hline

\end{tabular}}
\caption{$68 \% $ c.l. constraints on cosmological parameters in our extended $\Lambda$CDM+r+$N_{\rm eff}$ scenario from different combinations of datasets with a power-law inflation.}
\label{table2}
\end{center}
\end{table*}

We find the constraints on these 8 parameters by combining several recent datasets.
First of all, we call ``PlanckTT + lowTEB'' the full range of the 2015 temperature power spectrum ($2\leq\ell\leq2500$) combined with the polarization power spectra in the multipoles range $2\leq\ell\leq29$ provided by the \Planck collaboration \cite{Aghanim:2015xee}.
Secondly, when including the high multipoles \Planck polarization data \cite{Aghanim:2015xee}, we call this combination of datasets ``PlanckTTTEEE + lowTEB'' (which is considered less robust than the previous one by the \Planck collaboration, at least for order 1\,$\mu\mathrm{K}ˆ2$ wiggles relative to the plain \LCDM polarisation spectra). 
Afterwards, when replacing the lowTEB dataset with a gaussian prior on the reionization optical depth $\tau=0.055\pm0.009$, as obtained recently from \Planck HFI data \cite{newtau}, we refer to it as ``tau055''.
Moreover, we consider the 2015 \Planck measurements of the CMB lensing potential power spectrum $C^{\phi\phi}_\ell$ \cite{Ade:2015zua}, and we refer to this dataset as ``lensing''.
Finally, we add the CMB polarization $B$ modes constraints provided by the 2014 common analysis of \Planck, BICEP2 and Keck Array \cite{BKP}, and we refer to this dataset as ``BKP''.

For definiteness, we have used the June 2016 version of the publicly available Monte-Carlo Markov Chain package \texttt{cosmomc} \cite{Lewis:2002ah}, with a convergence diagnostic based on the Gelman and Rubin statistic. This version, which we modified to include the PLI case, implements an efficient sampling of the posterior distribution using the fast/slow parameter decorrelations \cite{Lewis:2013hha}, and it includes the support for the \Planck data release 2015 Likelihood Code \cite{Aghanim:2015xee} (see \url{http://cosmologist.info/cosmomc/}). 

\section{Results}

The result of these explorations are given in Tables~\ref{table1} and \ref{table2} where we report the constraints at $68 \% $ c.l. on the cosmological parameters, the two tables differing by considering or not the PLI specific constraint of eq.~\ref{eq:PLIdef}. By comparing the two Tables we see that imposing a PLI model affects the cosmological parameters in several ways. All the constraints that we will quote hereinafter there will be at $68\%$ c.l., unless otherwise expressed.

\begin{figure}[htbp]
\centering
\includegraphics[scale=1]{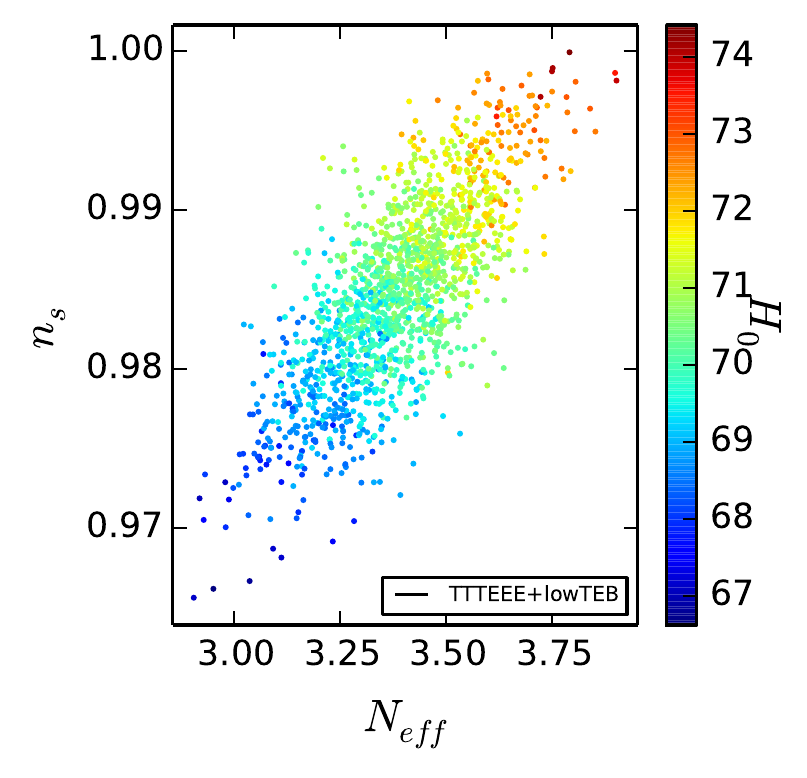}
\caption{Scattered 3D plot in the $\ns$ versus $N_{\rm eff}$ plane, coloured by $H_0$, under the assumption of PLI.}
\label{fig2}
\end{figure}

\begin{figure}[htbp]
\centering
\includegraphics[scale=1]{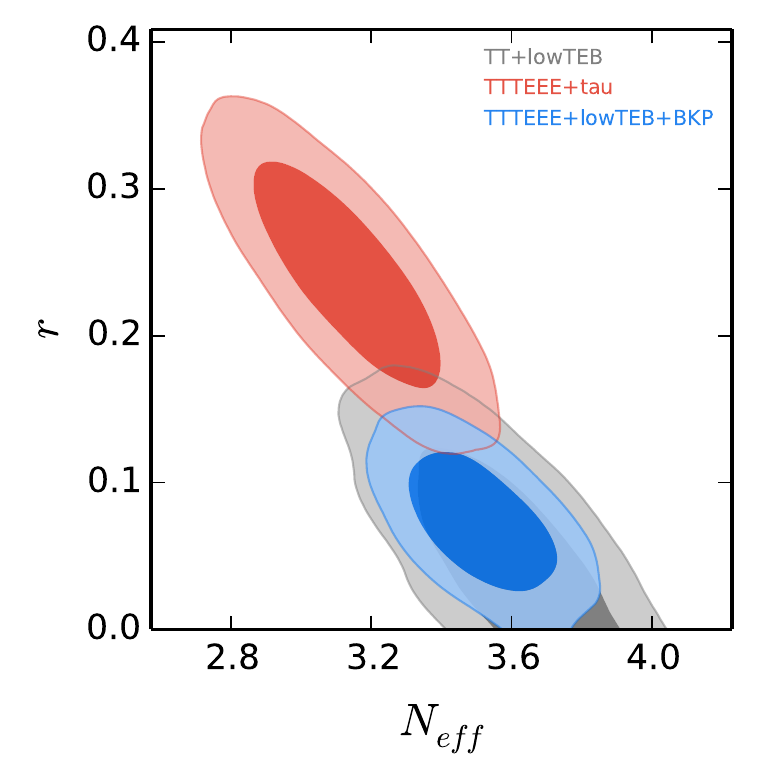}
\caption{Constraints at $68 \%$ and  $95 \%$ confidence levels on the $r$ vs $N_{\rm eff}$ plane under the assumption of PLI.}
\label{fig3}
\end{figure}

First of all, PLI produces a shift towards higher values of the neutrino effective number $N_{\rm eff}$ reducing the error by a half. For Planck TT + lowTEB, $N_{\rm eff}$ changes from $3.23_{-0.36}^{+0.30}$ to $N_{\rm eff}=3.55_{-0.17}^{+0.19}$. This shift corresponds to a higher $H_0$ value, $H_0=72.0_{-1.1}^{+1.5}$ Km/s/Mpc for the same combination of datasets, due to the degeneracy between these two parameters. This is illustrated in Fig.~\ref{fig2}. This increase of the Hubble constant parameter solves the tension existing between the local measurements provided by Riess et al. 2016 \cite{R16}, i.e., $H_0=73.00\pm1.75$ km/s/Mpc, and the value obtained from the \Planck CMB anisotropy data \cite{planck2015}, i.e. $H_0=67.27\pm0.66$ km/s/Mpc in a $\Lambda$CDM framework, as argued in \cite{Tram:2016rcw}. We note though that imposing the power-law inflation model, degrades somewhat the fit to the Planck TT + lowTEB data, producing a $\Delta \chi ^2 = 1.84$. 

\begin{figure}[htbp]
\centering
\includegraphics[scale=1]{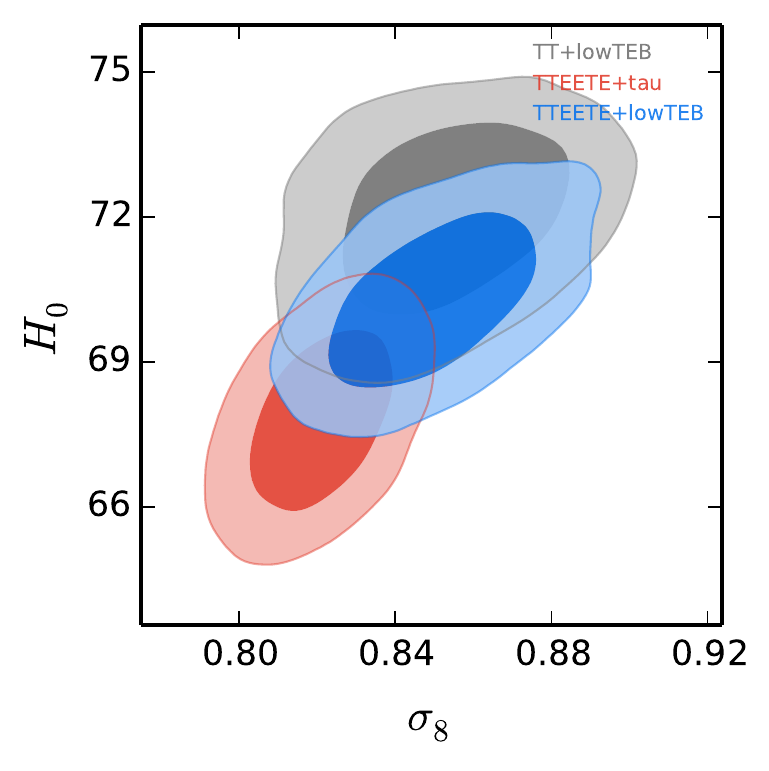}
\caption{Constraints at $68 \%$ and  $95 \%$ confidence levels on the $H_0$ vs $\sigma_8$ plane under the assumption of PLI.}
\label{fig4}
\end{figure}

\begin{figure}[bp]
\centering
\includegraphics[scale=1]{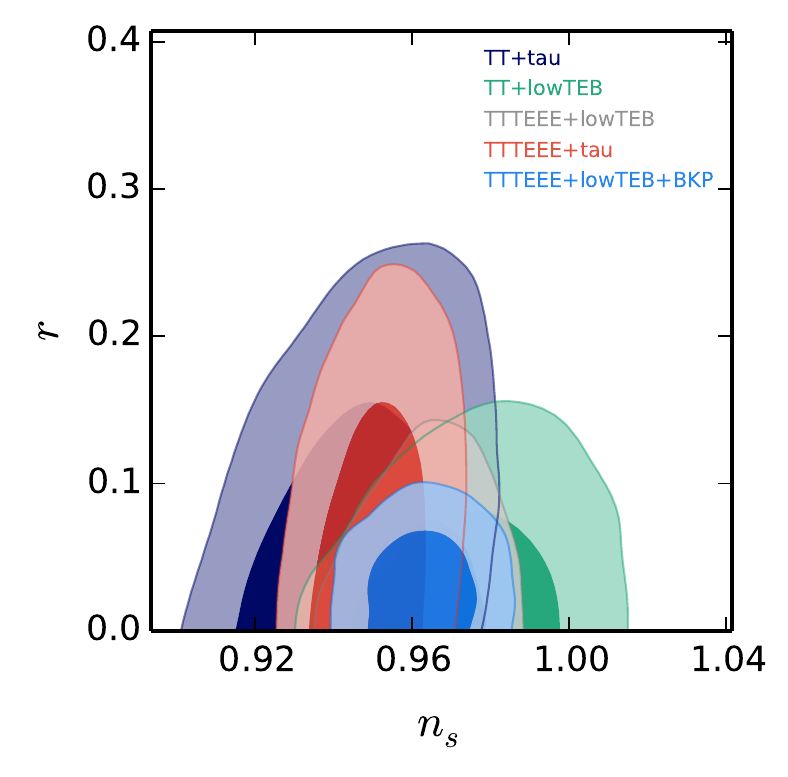}
\caption{Constraints at $68 \%$ and  $95 \%$ confidence levels on the $r$ vs $n_s$  plane, without considering the PLI.}
\label{fig1}
\end{figure}

Imposing PLI also leads to a higher value for the scalar spectral index \ns, that is only possible when $N_{\rm eff}$ is free to vary, since the two are positively correlated, as we can see in Fig.~\ref{fig2}. This is also the reason why \ns\ goes down again and produces evidence for a non zero tensor-to-scalar ratio, as we can see in Fig.~\ref{fig3}, when $N_{\rm eff}$ is further constrained by more data which shift it back towards the standard value. 

Indeed the new constraints on the reionization optical depth from \Planck HFI data \cite{newtau} reinstates compatibility with the standard value of $3.046$ for the neutrino effective number $N_{\rm eff}$ which restores the tension on $H_0$. Specifically, we find $N_{\rm eff}=3.24\pm0.22$ and $H_0=68.7\pm1.7$ Km/s/Mpc for Planck TT + tau055, and $N_{\rm eff}=3.14\pm0.17$ and $H_0=67.8\pm1.2$ Km/s/Mpc for Planck TTTEEE + tau055. Moreover, an evidence on $r$ different from zero at more than 3\,$\sigma$ appears: we find $r=0.209\pm0.064$ for Planck TT + tau055 and $r=0.240\pm0.048$ for Planck TTTEEE + tau055. In this case also, when imposing PLI, the fit to the Planck TT + tau055 data gets worse, producing a $\Delta \chi ^2 = 5.32$ for one less degree of freedom.

For the same reason, the clustering parameter $\sigma_8$ moves towards higher values when imposing PLI, since it is positively correlated with the Hubble constant, see Fig.~\ref{fig4}. Numerically, it moves from $\sigma_8=0.837_{-0.025}^{+0.022}$ to $\sigma_8=0.854\pm0.018$ when imposing a power-law inflation for Planck TT + lowTEB. This enhancement of the $\sigma_8$ value therefore increases notably the tension between the \Planck data and the weak lensing measurements from the CFHTLenS survey \cite{Heymans:2012gg, Erben:2012zw} and KiDS-450 \cite{Hildebrandt:2016iqg}. 

Finally, when we include the CMB polarization $B$ modes dataset provided by the common analysis of \Planck, BICEP2 and the Keck Array \cite{BKP}, a value of $r$ different from zero at more than 2$\sigma$ appears, i.e., $r=0.074_{-0.033}^{+0.027}$ for Planck TTTEEE + lowTEB + BKP in the power-law inflation, and the agreement between $H_0$ from Riess et al. 2016 \cite{R16} and the \Planck data is confirmed. Here again, the fit to the combination of datasets Planck TTTEEE + lowTEB + BKP gets worse  when imposing the power-law inflation, producing a $\Delta \chi ^2 = 9.14$ for one less degree of freedom. This evidence for $r$ different from zero is totally absent without imposing the power-law inflation, as we can see in Fig.~\ref{fig1}.

\section{Conclusions}
\label{sec:conclusions}

Recently, the authors of \cite{Tram:2016rcw} pointed out that by considering a power law inflation model and a free dark radiation component, the local measurements of the Hubble constant provided by Riess et al. 2016 \cite{R16}, i.e., $H_0=73.00\pm1.75$ km/s/Mpc at $68\%$ cl, is in perfect agreement with the value that follows when analysing the \Planck CMB anisotropy data \cite{planck2015}, inducing a $\Delta N_{\rm eff}=0.62\pm0.17$ at $68\%$ cl.

In this comment, we confront that scenario (PLI+dark radiation) with more data than initially considered. As noted in the previous section, the Hubble constant and the clustering parameter are positively correlated, therefore a higher $H_0$ value corresponds to an increased value of $\sigma_8$. When the $H_0$ tension subsides, this degeneracy produces a shift of the mean value of the clustering parameter which exacerbates the tension with the weak lensing measurements from the CFHTLenS survey \cite{Heymans:2012gg, Erben:2012zw} and KiDS-450 \cite{Hildebrandt:2016iqg} 

We then considered the implication of adding the CMB polarization $B$ modes dataset provided by the common analysis of \Planck, BICEP2 and Keck Array \cite{BKP}. In this case, interestingly, a value of $r$ different from zero at more than 2\,$\sigma$ is preferred, $r=0.074_{-0.033}^{+0.027}$ at $68\%$ cl for Planck TTTEEE + lowTEB + BKP (in the power-law inflation case), while the agreement between $H_0$ from Riess et al. 2016 \cite{R16} and the \Planck data is maintained.
However, when considering the new constraints on the reionization optical depth from Planck HFI data \cite{newtau}, the lower value preferred by that data on the neutrino effective number $N_{\rm eff}$ restores the tension on $H_0$ between the datasets and the standard value for $N_{\rm eff}$ is recovered.
The most recent data (given the $\Delta \chi ^2$) therefore does not really lend support to the power-law inflation model with dark radiation.

\section*{Acknowledgements}

\noindent We would like to thank S. Galli, T. Tram and V. Vennin for stimulating discussions. This work has been done within the Labex ILP (reference ANR-10-LABX-63) part of the Idex SUPER, and received financial state aid managed by the Agence Nationale de la Recherche, as part of the programme Investissements d'avenir under the reference ANR-11-IDEX-0004-02. 


\end{document}